\documentclass[a4paper,aps,11pt,showpacs,preprint]{revtex4}
\usepackage{graphicx,times,nicefrac,amsmath,amsfonts,amssymb,amsthm,epsf,bm,bbm,longtable,dcolumn}
\usepackage{revsymb,wasysym,mathrsfs,color}
\DeclareGraphicsExtensions{.pdf}
\begin{document}

\title{Main Magnetic Focus Ion Source: I. Basic principles and theoretical predictions}

\author{V. P. Ovsyannikov}\thanks{\url{http://mamfis.net/ovsyannikov.html}}
\affiliation{Hochschulstr. 13, D-01069  Dresden, Germany}

\author{A. V. Nefiodov}\thanks{Electronic mail: anef@thd.pnpi.spb.ru}
\affiliation{Petersburg Nuclear Physics Institute, 188300 Gatchina, St.~Petersburg, Russia}

\date{Received \today}
\widetext
\begin{abstract}

It is proposed to produce highly charged ions in the local potential traps formed by the rippled electron beam in a focusing magnetic field. In this method, the extremely high electron current densities can be attained on short length of the ion trap. The design the very compact ion sources is feasible. For such ions as, for example, Ne${}^{8+}$ and Xe${}^{44+}$, the intensities of about $10^9$ and $10^6$ particles per second, respectively, can be obtained.
\end{abstract}
\maketitle

\section{Introduction}

The ion traps formed by space-charge distribution of the electron beam together with the electrostatic potential barriers at some length were intensively investigated in the 40--60th years of the last century. The studies were caused by  attempts to develop the system of ion focusing for microwave devices \cite{1}. In the first experiments, ions were captured into the ion trap formed by the axially symmetric electron beam in the cylindrical drift tube with  positive potentials applied to both sides of the trapping region \cite{2}. The processes of compensation of the electron beam were considered theoretically in works \cite{2,3,4}, which treated the captured ions as an ideal gas obeying the Maxwell-Boltzmann distribution. However, the theory of the space-charge  compensation takes into account the singly charged ions only.

The concept of using the ion trap for production of \emph{multiply charged ions} was presented by E.D. Donets in 1967 \cite{5} following work by P.A. Redhead \cite{6}. The Donets's invention was named the electron-beam method for production of ions. In 1968, the first experimental device based on this method was demonstrated. The ions of gold with charges of up to $+19$ were successfully produced \cite{7}. The ion source received the name EBIS (electron-beam ion source). Afterwards, the ion sources have been widely used around the world for more than 40 years \cite{8}.

The specific applications of ion sources substantially affect their design (in particular, the length of ion trap $L_\mathrm{trap}$) and the operation regimes. The first generation of devices was aimed at the extraction of multicharged ions for subsequent employment at accelerators. In this case, the length of ion traps was of the order of $L_\mathrm{trap} \simeq 1$ m \cite{9,10,11,12}. In 1988, the ion source with $L_\mathrm{trap} \simeq 2$ cm in the trapping regime (without the ion extraction) received its own name EBIT (electron beam ion trap) was realized for the spectroscopical studies of the characteristic x-ray emission \cite{13}. The small length of ion trap allowed one to suppress the problem of plasma instabilities. Further breakthrough was associated with development of the ion sources without cryogenics \cite{14,15}. The peculiarity of ion sources of such type is to control the axial behavior of ions in the electron beam by using the \emph{external} electrostatic fields due to variation of potentials applied to different sections of the drift tube. Accordingly, the drift tube should consist of at least three sections.

Although presently there is a variety of devices with different names, all the modifications employ the same method of multiple sequential ionization by the electron beam suggested originally by E.D. Donets. The argumentation of this method can be applied for the electron beam with constant radius only and, therefore, with the smooth bottom of the axial potential distribution. In the following sections, we shall consider the rippled electron beam, which creates its own local ion traps. If the period of undulation is significantly less than the length of the single drift tube, these local ion traps cannot be controlled by the external electric fields in accordance with the Donets's scheme.

\section{Local ion traps in rippled electron beam}

\begin{figure}[hbtp]
\centerline{\resizebox{0.98 \columnwidth}{!}{ \includegraphics{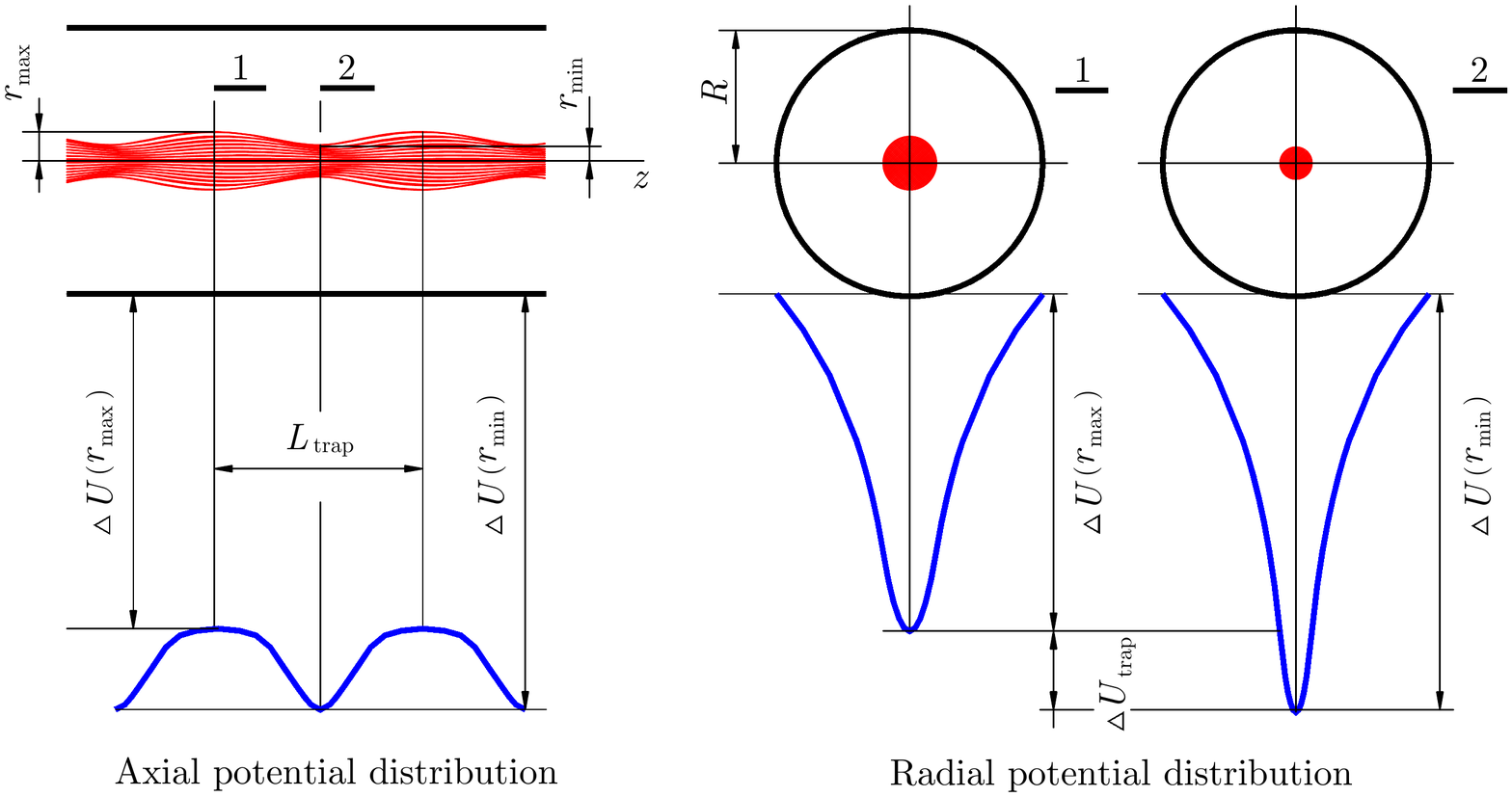}}}
\caption{\label{fig1} (Color online) Formation of local ion traps by the rippled electron beam, which propagates along the cylindrical drift tube. The electron beam is cut at the places, where the radius $r_e$ reaches either  the maximum $r_\mathrm{max}$ or the minimum $r_\mathrm{min}$.}
\end{figure}

The rippled electron beam with the radius $r_e$ varying periodically from the maximum $r_\mathrm{max}$ to the mi\-ni\-mum $r_\mathrm{min}$ creates a sequence of the local ion traps. The explanation of physical base for this type of ion traps is shown in Fig.~\ref{fig1}. The sag of the radial potential $\Delta U$ for the axially symmetric electron beam, which propagates along the drift tube with the radius $R$, is given by
\begin{equation}
\Delta U(r_e) =  \frac{U P}{4 \pi \varepsilon_0 \sqrt{2 \eta}} \left(1+2 \ln \frac{R}{r_e}\right) . \label{eq1}
\end{equation}
Here $\varepsilon_0$ is the permittivity of free space, $\eta = e/m$ is the magnitude of the electron charge-to-mass ratio, $U$ is the potential of the drift tube relative to the potential of the cathode, $P= I_{e}/U^{3/2}$ is the perveance of the electron beam, and $I_{e}$ is the electron current. Respectively, the depth of the axial potential well $\Delta U_\mathrm{trap}$ is equal to the difference between $\Delta U(r_\mathrm{min})$ and $\Delta U(r_\mathrm{max})$:
\begin{equation}
\Delta U_\mathrm{trap}= \Delta U(r_\mathrm{min})- \Delta U(r_\mathrm{max})= \frac{U P}{2 \pi \varepsilon_0 \sqrt{2 \eta}}  \ln \frac{r_\mathrm{max}}{r_\mathrm{min}}  . \label{eq2}
\end{equation}

As seen from Eq.~\eqref{eq2}, for the smooth electron beam ($r_\mathrm{max}=r_\mathrm{min}$), the local ion traps do not appear ($\Delta U_\mathrm{trap} = 0$). By contrast, for the rippled electron beam characterized, for example, by the ratio of radii $r_\mathrm{max}/r_\mathrm{min}=100$, the current $I_{e}$ of 1 A, and the energy $E_e=e U$ of 10 keV, the depth of the local potential well is equal to $\Delta U_\mathrm{trap}=1.4$ kV. Thus it is possible to create an effective local trap in the rippled electron beam without any external electric fields.

The length of the local ion trap $L_\mathrm{trap}$ along the $z$ axis is equal to half-length of the ripple wave. The value of $L_\mathrm{trap}$ depends on the accelerating voltage $U$, the distribution of the focusing magnetic field $B(z)$, and the parameter of cathode conditions $K$. The parameter $K$ is defined as the ratio of the magnetic flux through the cathode with the radius $r_c$ to the magnetic flux through the cross section of the electron beam with the radius $r_e$. This parameter is very important for the paraxial theory. Whereas $K$ is varied between $0$ and $1$, the electron trajectories are changed significantly. The value $K=0$ corresponds to so-called the ``Brillouin's focusing system" \cite{16}. In this case, the cathode of the electron gun should be located in the zero magnetic field. Theoretically, for given magnetic field and the energy of electron beam, the extended Brillouin's beam with constant radius acquires the highest current density \cite{17}. This property of the Brillouin's focusing makes it very attractive for EBIS. However, the practical realization of such electron beam focusing is rather difficult \cite{18}. Extremely careful shapes both for the magnetic field and for the electron trajectories in electron gun are necessary to obtain the nonrippled beam. There is only one combination of the shape of magnetic field and the electron gun geometry, for which the electron beam can have the constant Brillouin's radius. These are so-called the ``Brillouin's conditions" \cite{16}. In all other cases, the electron beam is rippled.

Generally, for arbitrary relationship between the magnetic field distribution and the electron trajectories, the thermal theory predicts formation of a sequence of images and crossovers \cite{19}. In the crossovers, that is, in the local ion traps created by the rippled beam, the electron current density is many times higher than that in the case of the smooth Brillouin's flow. Therefore, new generation of the ion sources characterized by both small size and extremely high electron current density can be realized by using local ion traps in the rippled electron beam. A low capacity of the ion trap due to very short length of the device can be compensated by running in the mode with high repetition rate.

\section{Formation of local ion traps with extremely high electron current density}

In general case, the focusing of the axial electron beam in the magnetic field of a solenoid is the focusing of charged particles by a thick magnetic lens. The electron beam from the cathode at zero magnetic field is transformed into a sequence of the focuses in the regular magnetic field. Since the first focus is the most acute, it can be called the \emph{main} magnetic focus. In the next focuses, the diameter of the beam increases successively under the influence of aberrations and nonlinear effects. A magnetically compressed electron beam was studied by K. Amboss \cite{19}. The main conclusion of this work is that the electron beam damps out the undulations and looses the periodic structure within a relatively short distance. Therefore, in the following, we shall take into account the magnetic lens for three focuses only.

\begin{figure}[bthp]
\centerline{\resizebox{0.8 \columnwidth}{!}{ \includegraphics{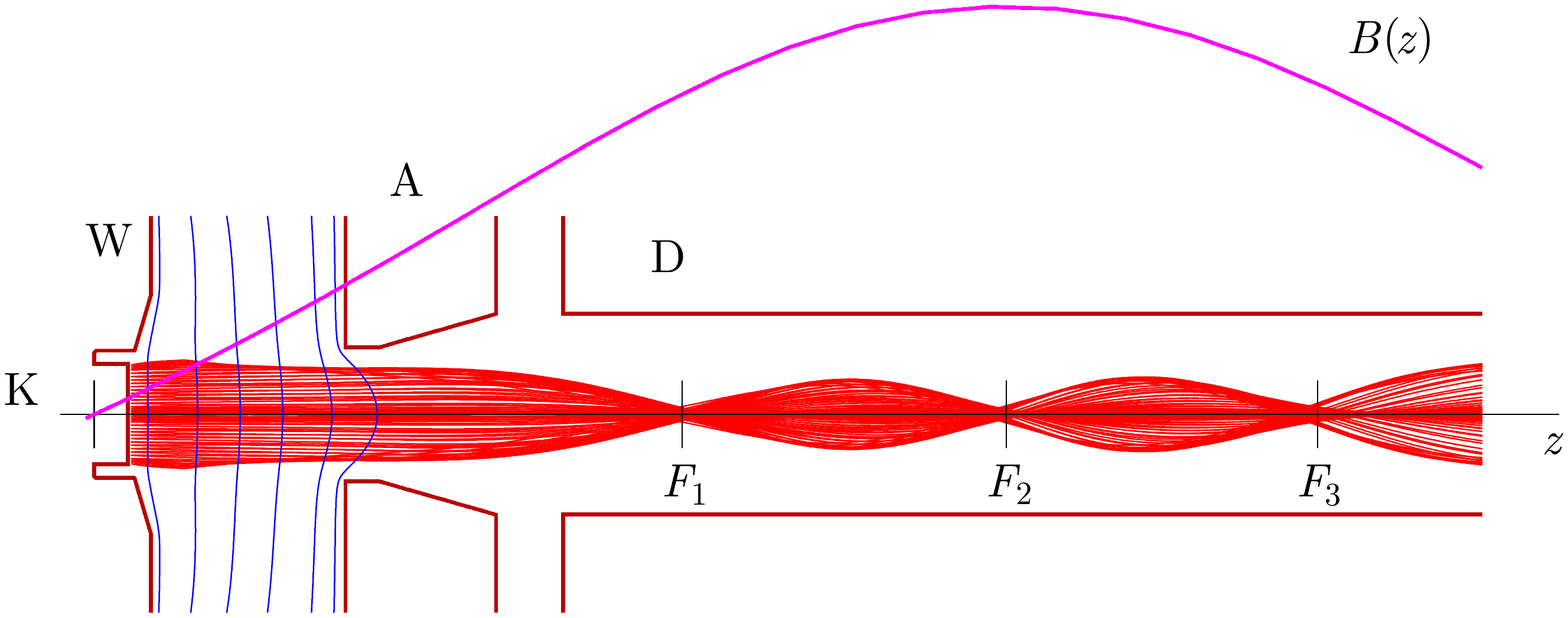}}}
\caption{\label{fig2} (Color online) Focusing of the electron beam by a thick magnetic lens. The electron gun is arranged with cathode (K), focusing (Wehnelt's) electrode (W), and anode (A). The drift tube is marked by the letter D, while $F_1$, $F_2$, and $F_3$ denote the first, second, and third focuses, respectively. }
\end{figure}

\begin{figure}[hbtp]
\centerline{\resizebox{0.45 \columnwidth}{!}{ \includegraphics{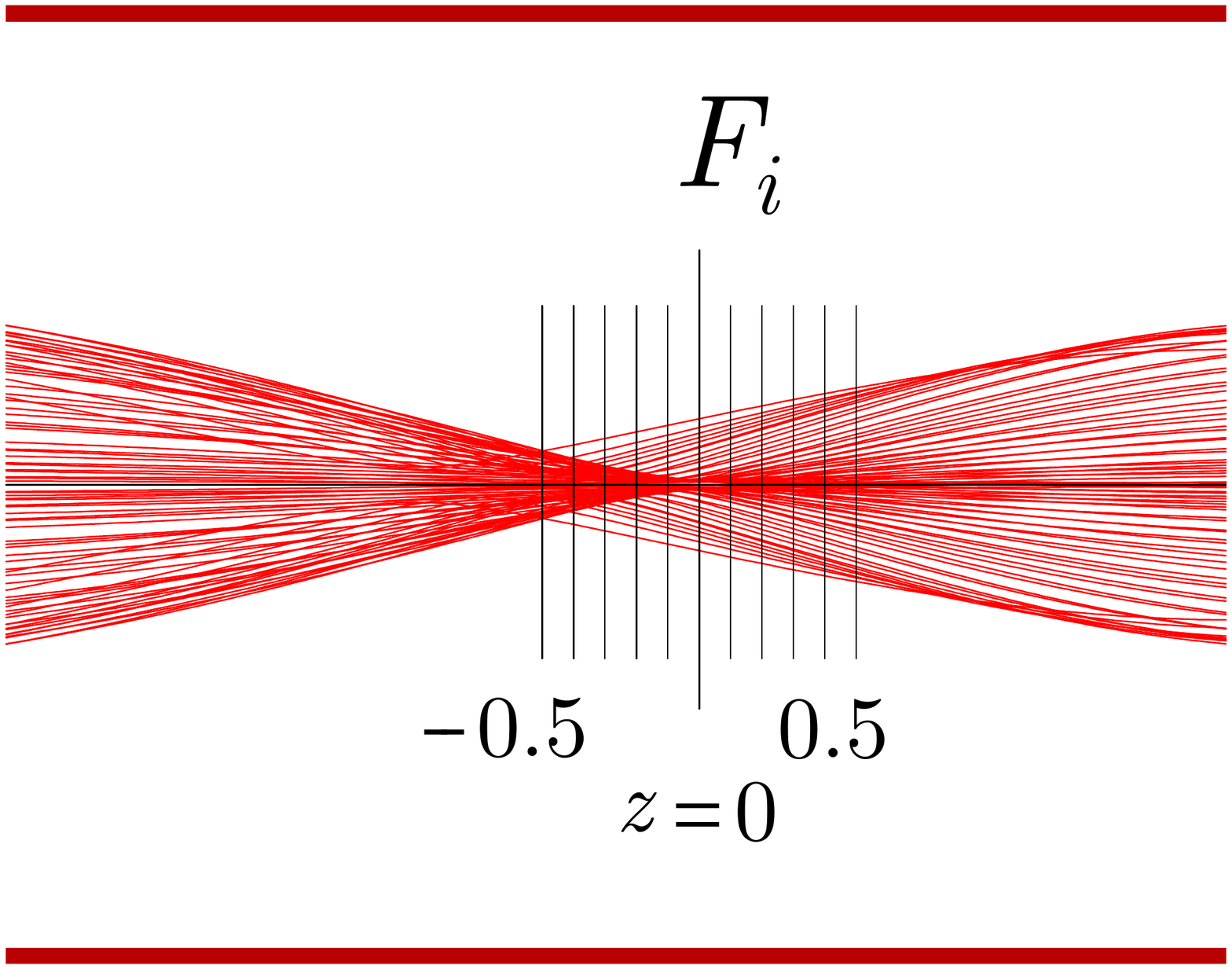}}}
\caption{\label{fig3} (Color online) Overview of the magnetic focus $F_i$ and the positions of cuts, where the electron current densities are calculated.}
\end{figure}

As an example, let us consider the electron beam with the current $I_e= 50$ mA and the energy $E_e= 10$ keV projected by cathode with the radius $r_e=0.25$ mm. The magnetic field has the short distribution $B(z)$, which allows to create three focuses. The electron trajectories in the focusing magnetic field are shown in Fig.~\ref{fig2}. In each focus, the local ion trap appears. The axial distributions of the electron current density are calculated over the region of each focus, namely, from the point $z = -0.5$ mm to the point $z = +0.5$ mm with the step of $0.1$ mm (see Fig.~\ref{fig3}). The central point $z=0$ corresponds to the position of the highest current density in each focus.

\begin{figure}[bthp]
\centerline{\resizebox{0.6 \columnwidth}{!}{ \includegraphics{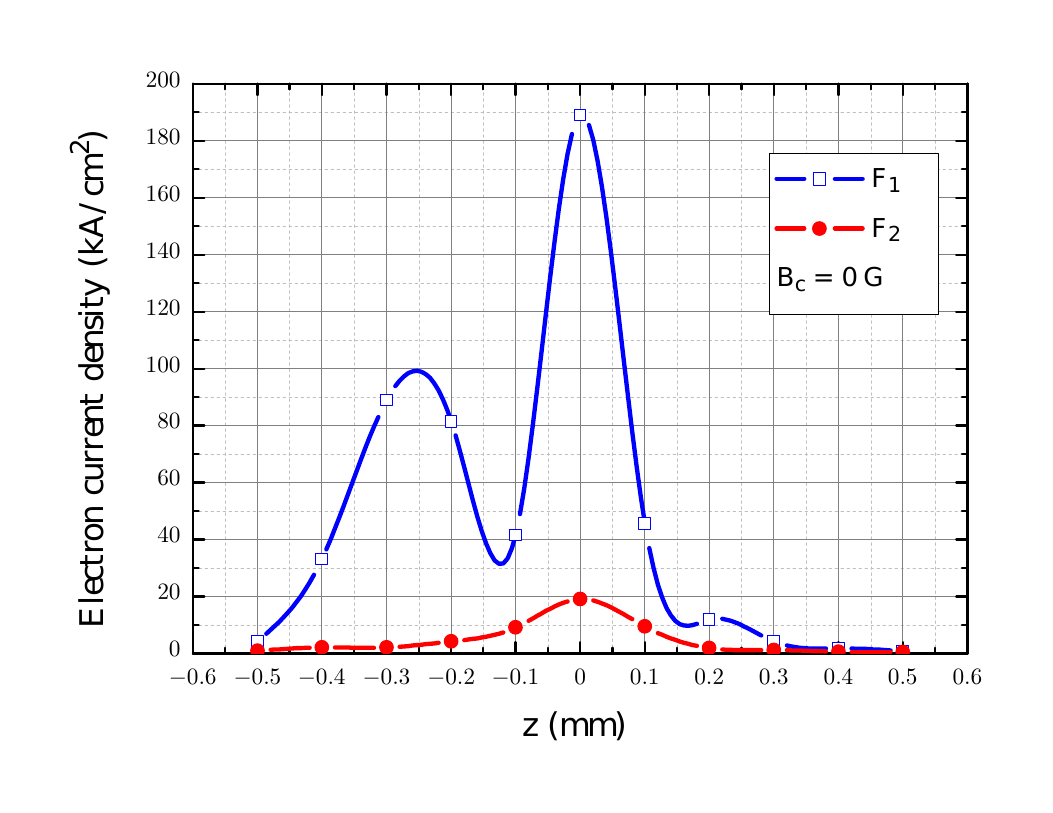}}}
\caption{\label{fig4} (Color online) Axial distribution of the electron current density within the local ion trap for the first and second focuses. The magnetic field at the cathode is absent. The point $z =0$ is adjusted to the position of the highest current density in each focus.}
\end{figure}

\begin{figure}[bthp]
\centerline{\resizebox{0.6 \columnwidth}{!}{ \includegraphics{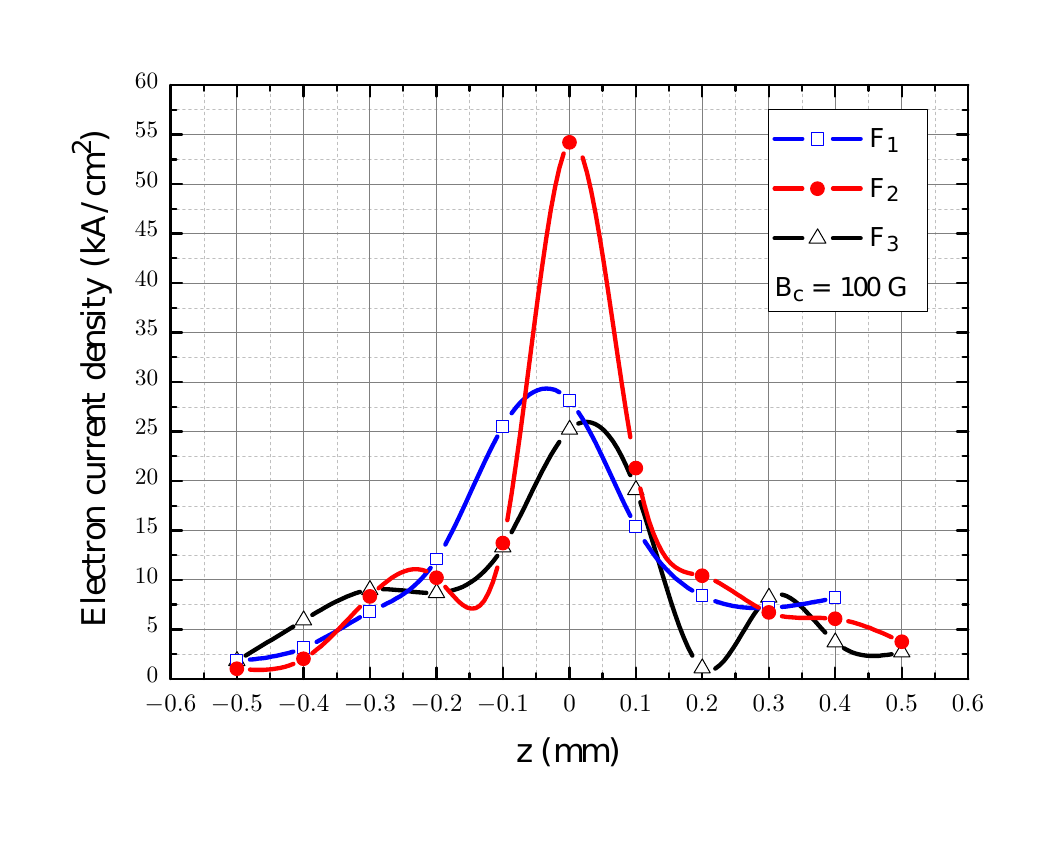}}}
\caption{\label{fig5} (Color online) Distribution of the electron current density for the case of $B_c = 100$ G.}
\end{figure}

\begin{figure}[bthp]
\centerline{\resizebox{0.6 \columnwidth}{!}{ \includegraphics{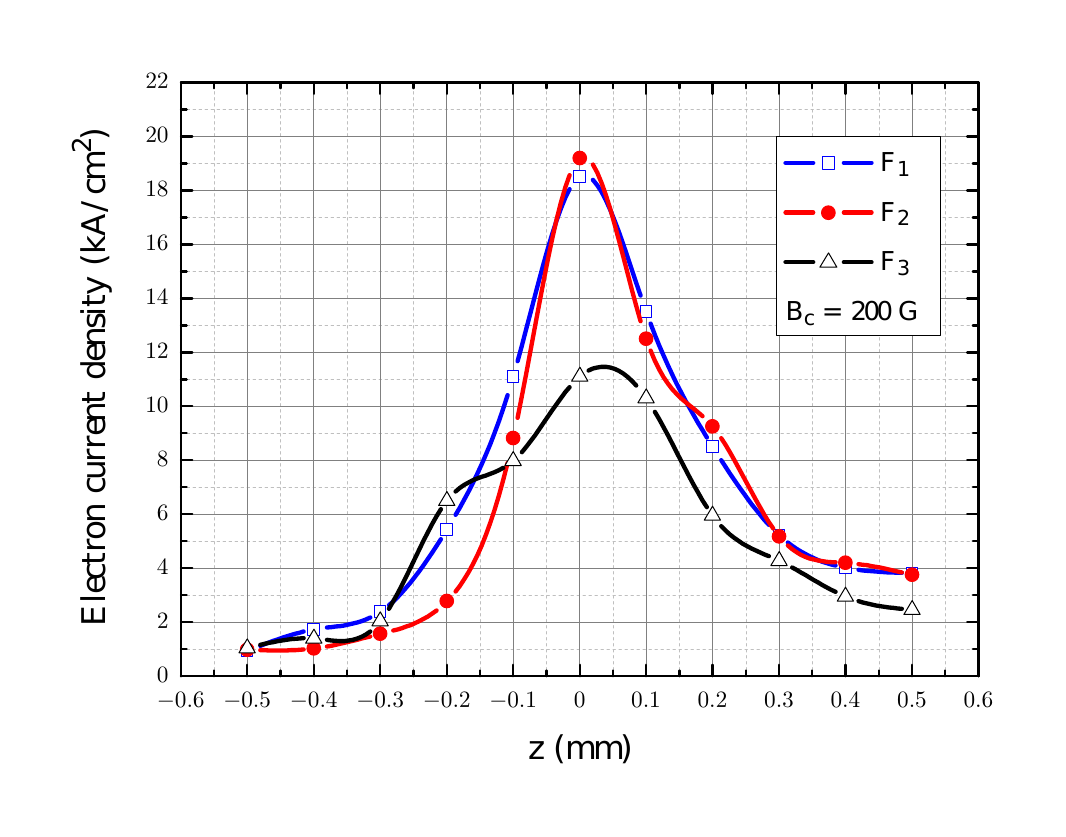}}}
\caption{\label{fig6} (Color online) Similar to Fig.~\ref{fig5} for the case of $B_c = 200$ G.}
\end{figure}

The results of numerical calculations of the electron current densities are presented in Figs.~\ref{fig4}~--~\ref{fig6} for different magnetic fields at the cathode $B_c$. For the zero field at the cathode, the first focus is acute (see Fig.~\ref{fig4}), as it is predicted by the theory of the thick magnetic lenses. The structure of this focus is complicated. It exhibits all general features for focusing of the electrons emitted by the cathode with finite size. One can distinguish two groups of the electrons, which create two peaks, respectively. One part of the electrons is ejected from the central part of the cathode and generates the more acute peak. The second group of the electrons is emitted from the peripheral part of the cathode. The electrons are subjected to greater influence of the thermal velocity effects, aberrations of anode lens, and the distortion effects of the electric field in the gap between the cathode and focusing electrode. These effects give rise to the turbulence of the electron beam just in the first focus. The errors generated by the parasitic processes are accumulated very rapidly, so that the subsequent focuses lose their densities. The amplitude value of the electron current density in the second focus is less than that in the first focus by about factor of 10. The density distribution in the third focus is negligibly small. Nevertheless, it is possible to achieve very high density in the second focus of the magnetic lens, if the defocusing processes are suppressed.

The magnetic field at the cathode $(K>0)$ significantly alters the electron trajectories. Just a slight magnetic field at the cathode suppresses some defocusing factors, namely, the effect of thermal velocities of electrons and distortion of the electric field near the boundary of the cathode. However, these positive effects cause also some losses of the electron current density.

In Fig.~\ref{fig5}, the distributions of the electron current density are shown for three focuses of the electron beam, if $B_c = 100$ G. The maxima of the current density are decreased. The strongest peak is now located in the second focus $F_2$. The electron current density along the $z$ axis is more close to the normal distribution. Despite of some losses of the electron density, the amplitude value of the current density in the second focus is high enough ($\sim 55$~kA/cm$^2$).

Increasing magnetic field at the cathode up to $B_c = 200$ G leads to further reduction of the maximum electron current density in the focuses of the electron beam (see Fig.~\ref{fig6}). In the range of each focus, the electron current density becomes more close to the normal distribution. The current density turns out to be more than 10 kA/cm$^2$ on the length of about 0.3 mm. The values of the current density in different focuses are close to each other.

The examples considered above show that it is possible to design the miniature device, in which the geometric factor of the rippled electron beam is used to confine the ions in the local ion traps. The electron current density can achieve extremely high values of the order of hundred kA/cm$^2$ on the length of, at most, 1 mm. We shall refer to the device as the main magnetic focus ion trap (MaMFIT), in accordance with definition of the most acute focus as the main one.

The following two remarks should be pointed out here. Firstly, the present consideration is performed for pure electron beam. Since the ions prepared in the local ion traps compensate the space charge of electrons, the real electron current densities should be higher than those predicted in Figs.~\ref{fig4}--\ref{fig6}. The ion focusing of the electron beam needs to be included in the model calculations. Secondly, further increase of the magnetic field at the cathode leads to the system with magnetic compression of the electron beam. The main estimates for such system were done for the EBIS KRION-2 in Dubna \cite{12}. This method of focusing makes it possible to obtain  relatively smooth electron beam with high current density \cite{20}.

\section{Control over behavior of ions in local ion traps}

The ion trap together with the system of ion extraction becomes the ion source. In particular, the MaMFIT is transformed into the MaMFIS (main magnetic focus ion source). However, the standard method used for control over behavior of ions in the ion trap of the EBIS, where the axial transport of ions is managed by the external electrostatic fields, is not feasible for the MaMFIS. The implementation of this principle would require to locate the drift tubes very close to each other. As a result, the interelectrode space becomes too small initiating the voltage breakdown. Since the electron beam makes the local ion traps by its rippled structure, the electron beam should change its form into the smooth one in order to allow the ion extraction. In this case, the potential distribution along the electron beam becomes also smooth, so that the ions can leave the local trap. Practically, it can be realized, if the potential of the focusing electrode is switched over negative with respect to that of the cathode. Then the pulsed electron beam is transformed into the smooth beam. Although this method requires special shapes both for the electrodes of the electron gun and for the magnetic field, it allows the MaMFIS to run with high repetition rate. A very short time requisite for the ion confinement is provided by extremely high electron current density in the local ion trap.

\begin{figure}[hbt]
\centerline{\resizebox{0.9 \columnwidth}{!}{ \includegraphics{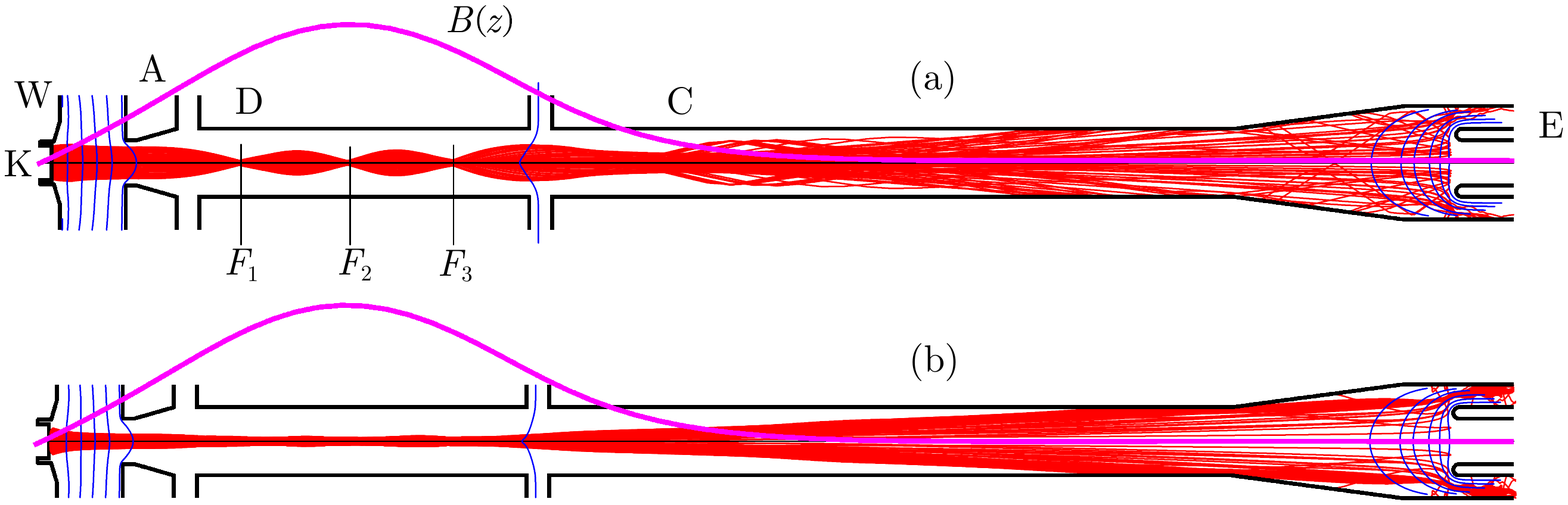}}}
\caption{\label{fig7}  (Color online) Schemes of the electron optics and the electron trajectories for the MaMFIS with three focuses. Operation regimes: (a) trapping mode and (b) extraction mode. The marks are similar to those employed in Fig.~\ref{fig2}. In addition, the letter C denotes the collector and E denotes the extractor (ion optics). }
\end{figure}

In Fig.~\ref{fig7}, the schematic structure of the three-focus MaMFIS, which corresponds to calculations presented in Figs.~\ref{fig3}--\ref{fig6}, is shown. The electron beam has the current $I_e=0.1$ A and the energy of $E_e=10$ keV. The amplitude value of the magnetic field strength is about 4 kG. The trapping mode is realized, if the cathode potential is equal to the potential of the focusing electrode. When the potential of the Wehnelt's electrode is less than the cathode potential, the electron beam becomes smooth and the extraction mode is switched on.

\section{Production of highly charged ions in MaMFIS}

The rough estimate shows that the confinement time for the production of highly charged ions in the electron beam with extremely high current density is sufficiently small \cite{13}. In addition, the length of the local ion trap is very short. Accordingly, injection of the working substance into the ion trap as a pulse in accordance with basics of the EBIS technology is either difficult or impossible. In particular, it concerns light elements of the Periodic Table.

We shall consider the preparation of high charge states in the local ion trap with \emph{permanent} pressure of the working substance. The standard EBIS theory  predicts that in this case the output spectrum should contain all ions of the working gas with the charges $q$ from $+1$ up to some maximum value \cite{20}. However, in the plasma trapped by extremely dense electron beam, the phenomenon of ``auto-cooling" takes place. Namely, the low charged ions of the working substance cool the highly charged ions of the same working substance. In other words, there is no need to add any special cooling gas, since the working gas cools itself. As a result, the equilibrium spectrum of ions turns out to be very narrow. It consists of highly charged ions mainly. This phenomenon is realized, because the neutral atoms and the ions with low charge states of about $+1$, which are generated continuously from the working substance, cool the highly charged ions accumulating in the trap. The ions in the middle charge states escape rapidly from the trap due to the process of ``evaporative cooling" \cite{21}.

If the energy of incoming electrons is fixed, there is the limiting concentration of neutral atoms $N_0^\mathrm{lim}$, in which the production of ions with the given charge state $q$ is still possible. The limiting concentration $N_0^\mathrm{lim}$ can be estimated from the condition of dynamical equilibrium between the electron loss and capture processes. If the radiative recombination is less probable than the charge exchange of multicharged ions on neutral atoms, the balance of rates for the ionization and the charge exchange is equivalent to the following equation:
\begin{equation}\label{eq3}
N_0^\mathrm{lim} = \frac{j_e}{e} \frac{N_{q-1} \sigma^+_{q-1, q}}{N_q v \sigma^\mathrm{chex}_{q, q-1}}  .
\end{equation}
Here $N_q$ is the concentration of ions in the charge state $q$, $v$ is the ion velocity averaged over the Maxwellian distribution, $\sigma^\mathrm{chex}_{q, q-1}$ is the cross section for charge exchange of ions with neutrals, and $\sigma^+_{q-1, q}$ is the single ionization cross section. Complete consideration of this running mode will be presented in details elsewhere. As it follows from Eq.~\eqref{eq3}, the limiting concentration is directly proportional to the electron current density $j_e$. Therefore, in the local ion trap formed by very dense electron beam, highly charged ions can be produced in low vacuum.

In order to estimate the operation possibilities of the MaMFIS, we have studied the charge state distributions and expected intensities for ions of Xe, Ar, and Ne produced in the local ion trap by the electron beam with the current density $j_e=20$ kA/cm$^2$. The calculations are performed with the use of the computer code written in 1996-1999 years \cite{22,23}. The code takes into account two working substances with permanent pressures in the ion trap. One of the substances can be chosen as a residual gas, for example, hydrogen. This option allows one to simulate the ionization processes most closely to the experimental conditions. The parameters of the local ion trap used in the numerical calculations are presented in Table~\ref{tab1}.

\begin{table}[tbh]
\caption{\label{tab1} MaMFIS: parameters of ion trap}
\begin{center}
\vspace{0.2 cm}
\begin{tabular}
{ l c|c l}
\hline  \hline
Electron current $I_e$    &&&   0.1 A \\
Electron energy $E_e$  &&&  10 keV \\
Electron current density $j_e$  &&&  20 kA/cm$^2$ \\
Length of ion trap $L_\mathrm{trap}$  &&&  $0.1$ cm \\
Volume of ion trap $V_\mathrm{trap}$  &&&  $5 \times 10^{-7}$ cm$^3$\\
Total electron charge $Q_e$  &&&  $10^{7}$ $e$ \\
Concentration of electrons $n_e$  &&&  $2 \times 10^{13}$ cm$^{-3}$ \\
\hline  \hline
\end{tabular}
\end{center}
\end{table}

\subsection{Ionization of Xe}

\begin{figure}[tbhp]
\centerline{\resizebox{0.8 \columnwidth}{!}{ \includegraphics{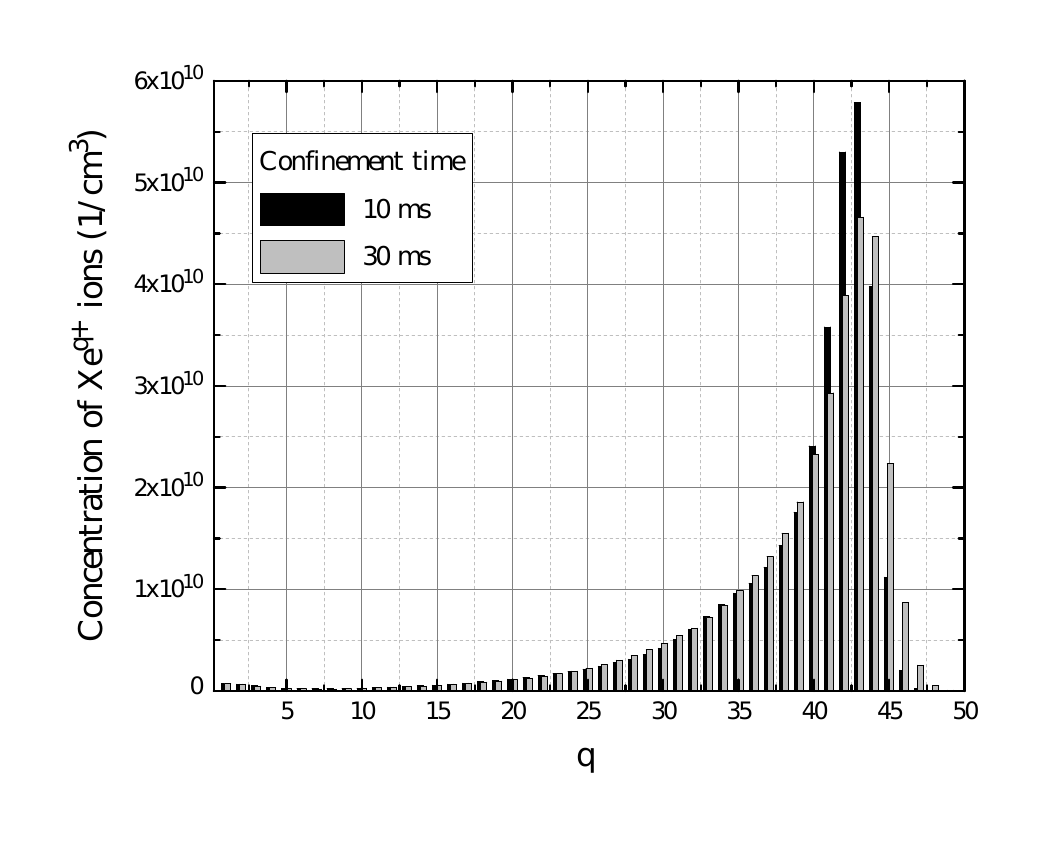}}}
\caption{\label{fig8} Charge state distributions of Xe$^{q+}$ ions at permanent pressure of the working gas.}
\end{figure}

The simulation is performed for the permanent pressure of neutral Xe of $2 \times 10^{-8}$ mbar with admixture of neutral hydrogen under the partial pressure of $10^{-9}$ mbar (see Fig.~\ref{fig8}). The confinement time is chosen to be 10 and 30 ms. Accordingly, the MaMFIS should run with the repetition rate $\nu$ of about $100$ -- $35$ Hz. The spectrum is characterized by considerable suppression of the ion population in the middle charge states within the range $+5 \lesssim  q \lesssim +15$, which actively participates in the process of ``evaporative cooling". In addition, the part of highly charged ions is pronounced and relatively narrow.

The yield of ions with charge $q$ can be estimated by using the following formula:
\begin{equation}\label{eq4}
\Upsilon =  \nu N_q V_\mathrm{trap} .
\end{equation}
Taking the concentration of Xe$^{44+}$ ions $N_q$ of about $4 \times 10^{10}$ cm$^{-3}$, the ionization volume $V_\mathrm{trap}= 5 \times 10^{-7}$ cm$^3$, and the running frequency $\nu = 100$ Hz, one obtains the ion yield $\Upsilon \simeq 2 \times 10^6$ particles per second (pps). This value is much higher than the intensities obtained with the use of the Dresden EBIS devices \cite{24}.

\subsection{Ionization of Ar}

\begin{figure}[hbtp]
\centerline{\resizebox{0.8 \columnwidth}{!}{ \includegraphics{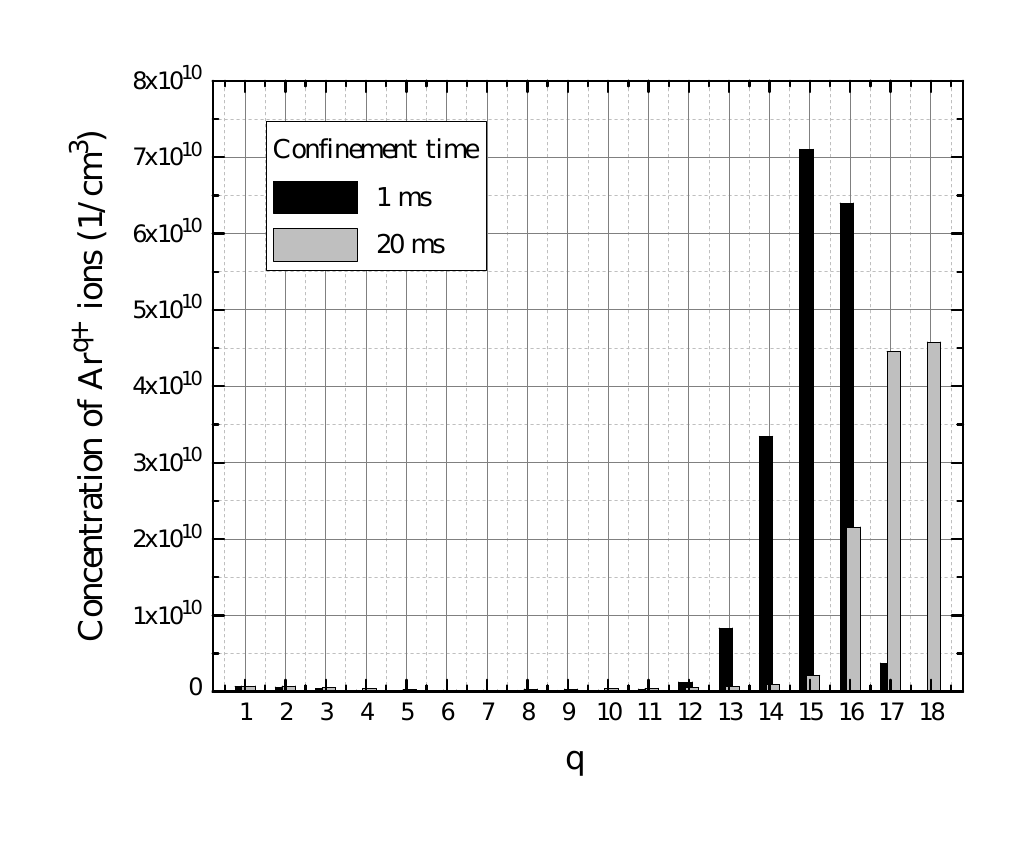}}}
\caption{\label{fig9} Charge state distributions of Ar$^{q+}$ ions. The permanent pressures of neutral gases are equal to $2 \times 10^{-8}$ mbar and $10^{-8}$ mbar for argon and hydrogen, respectively. The parameters of the electron beam are the following: $I_e=0.1$ A, $j_e=20$ kA/cm$^2$, and $E_e=8.5$ keV.}
\end{figure}

The charge spectra of Ar$^{q+}$ ions after their confinement in the local ion trap during 1 and 20 ms are depicted in Fig.~\ref{fig9}. The estimates for the ion current are made analogously to that for the ions of Xe. For example, after 1 ms the predicted yield of Ar$^{16+}$ ions should be about $2 \times 10^{7}$ pps. For the production of Ar$^{18+}$ ions the confinement time is increased up to 20 ms. The corresponding estimate for the ion yield gives $\Upsilon \simeq 10^{6}$ pps.

\subsection{Ionization of Ne}

\begin{figure}[hbtp]
\centerline{\resizebox{0.8 \columnwidth}{!}{ \includegraphics{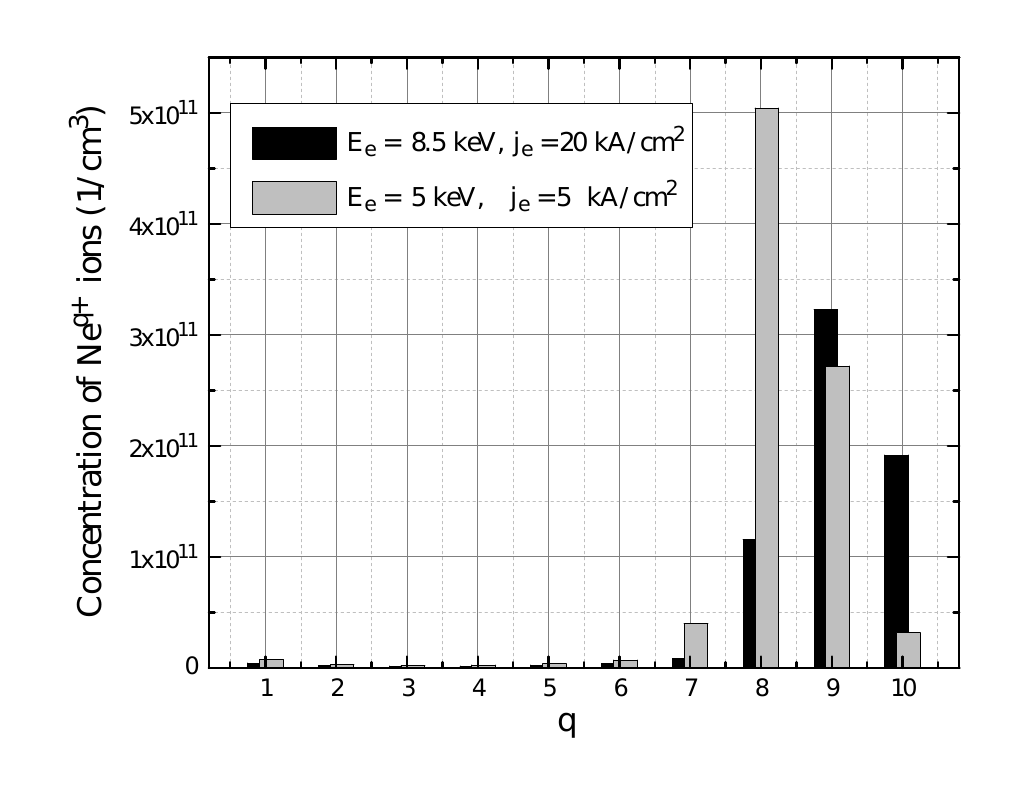}}}
\caption{\label{fig10} Charge state distributions of Ne$^{q+}$ ions. The confinement time is 1 ms. Neutral hydrogen under the permanent partial pressure of $10^{-8}$ mbar is admixed as the background gas.}
\end{figure}

For stepwise ionization of the light elements such as Ne, the pressure of neutral gas in the ion trap can be increased. In the first computer simulation, the permanent pressure of Ne is fixed at the level of $2 \times 10^{-7}$ mbar. The admixture of residual hydrogen is also taken into account under the permanent pressure of $10^{-8}$ mbar. The electron beam is characterized by $I_e=0.1$ A, $j_e=20$ kA/cm$^2$, and $E_e=8.5$ keV. The charge spectrum of Ne ions after 1 ms of confinement is presented in Fig.~\ref{fig10}. The ion yield of Ne$^{9+}$ is estimated as about $2 \times 10^8$ pps, while the yield of Ne$^{10+}$ ions is less by half.

In order to compare the influence of electron current density and vacuum on the charge spectrum of Ne ions, the second simulation is done for the electron beam with $j_e=5$ kA/cm$^2$ and $E_e=5$ keV in vacuum of $10^{-6}$ mbar. In this case, the ionization volume $V_\mathrm{trap}$ is of about $2 \times 10^{-6}$ cm$^3$. The background gas (hydrogen) is assumed to have the permanent pressure of $10^{-8}$ mbar. The charge state distribution of Ne ions is shown in Fig.~\ref{fig10}. The confinement time is 1 ms. The yield of Ne$^{8+}$ is estimated to be at least $10^9$ pps. Obviously, the ions can be prepared for the confinement time even less than 1 ms.

The theoretical estimates presented above confirm the well-known statement that the high electron current density is the key for effective production of highly charged ions. The pilot examples of the MaMFIS/T with the electron beam energies of 4, 10, and 30 keV are being tested by the method of x-ray spectroscopy \cite{25}. The experimental results will be published elsewhere.

\section{Conclusions}

The technology of magnetic compression of the electron beam from the cathodes with high emission density allows one to create the unique ion traps in crossovers of the rippled electron beam. The control over behavior of ions can be performed by changing the potential on the focusing electrode of the electron gun. The ion traps have the extremely high electron current density in crossovers and very short length. The latter gives one more advantage of the device, because it decreases the chance of instability of the dense electron beams. The MaMFIS is the compact ion source of new generation, which can be used for technological applications and scientific investigations in atomic physics, plasma physics, solid-state physics, single ion implantation, ion lithography, and others.

\section*{ACKNOWLEDGMENTS}

The authors are grateful to O. K. Kultashev for his contribution to creation of the electronic optics and to Aleksander A. Levin for his support.


\begin{thebibliography}{99}
\bibitem{1} J. R. Pierce, \emph{Theory and design of electron beams} (Van Nostrand, New-York, 1954).
\bibitem{2} L. M. Field, K. Spangenberg, and R. Helm, Elec. Communication {\bf 24}, 108 (1947).
\bibitem{3} O. Scherzer, Z. Physik  {\bf 82}, 697 (1933).
\bibitem{4} E. G. Linder and K. G. Herngvist, J. Appl. Phys. {\bf 21}, 1088 (1950).
\bibitem{5} E. D. Donets, USSR Inventor's Certificate No. 248860, 16 March (1967);
Bull. OIPOTZ {\bf  23}, 65 (1969).
\bibitem{6} P. A. Redhead, Can. J. Phys. {\bf 45}, 1791 (1967).
\bibitem{7} E. D. Donets, V. I. Iluschenko, and V. A. Alpert, \emph{Electron-beam source of multicharged ions}, JINR Communication R7-4124, Dubna (1968).
\bibitem{8} R. Becker and O. Kester, Rev. Sci. Instrum. {\bf 81}, 02A513 (2010).
\bibitem{9} V. G. Aksenov, E. D. Donets, A. G. Zeldovich, A. I. Pikin, and
Yu. A. Schischov, \emph{Cryogen-magnetic system for electron-beam source of multicharged ions (``KRION")},
JINR Communication R8-8563, Dubna (1975).
\bibitem{10} E. D. Donets and A. I. Pikin, Zh. Tekh. Fiz. {\bf 45}, 2373 (1975)[Sov. Phys. Tech. Phys.  {\bf 20}, 1477 (1976)].
\bibitem{11} J. Arianer and C. Goldstein, IEEE Trans. Nucl. Sci. {\bf 23}, 979 (1976).
\bibitem{12} E. D. Donets and V. P. Ovsyannikov, \emph{Cryogenic electron-beam ion source ``KRION-2"}, JINR Communication R7-9799, Dubna (1976).
\bibitem{13} M. A. Levine, R. E. Marrs, J. R. Henderson, D. A. Knapp, and M. B. Schneider,
Phys. Scr.  T{\bf 22}, 157 (1988).
\bibitem{14} H. Khodja and J. P. Briand, Phys. Scr. T{\bf 71}, 113 (1997).
\bibitem{15} V. P. Ovsyannikov and G. Zschornack, Rev. Sci. Instrum.  {\bf 70}, 2646 (1999).
\bibitem{16} G. Hermann, J. Appl. Phys.  {\bf 29}, 127 (1958).
\bibitem{17} P. T. Kirstein, G. S. Kino, and W. E. Waters, \emph{Space-Charge Flow}
(McGraw-Hill, New-York, 1967), Ch. 8.
\bibitem{18} J. Arianer, A. Gabrespine, C. Goldstein, T. Junquera, A. Courtois, G. Deschamps, and  M. Oliver,
Nucl. Instrum. Meth. {\bf 198}, 175 (1982).
\bibitem{19} K. Amboss, IEEE Trans. Electr. Dev. {\bf  16}, 897 (1969).
\bibitem{20} R. Becker, Rev. Sci. Instrum.  {\bf 71}, 816 (2000).
\bibitem{21} B. M. Penetrante, J. N. Bardsley, M. A. Levine, D. A. Knapp, and R. E. Marrs, Phys. Rev. A {\bf 43}, 4873 (1991).
\bibitem{22} I. V. Kalagin and V. P. Ovsyannikov, \emph{Numerical simulation of ion production processes in EBIS}, JINR  Communication E9-96-128, Dubna (1996).
\bibitem{23} I. V. Kalagin, D. K\"{u}chler, V. P. Ovsyannikov, and G. Zschornack, Plasma Sources Sci. Technol. {\bf 7}, 441 (1998).
\bibitem{24} V. P. Ovsyannikov and G. Zschornack, J. Instrum.  {\bf 5}, C11002 (2010).
\bibitem{25} V. P. Ovsyannikov, \emph{Main Magnetic Focus Ion Trap, new tool for trapping of highly charged ions},    arXiv e-print (arXiv:1403.2168) (2014).
\end{thebibliography}
\end{document}